%% file: main.tex
\documentclass[3p]{elsarticle}

\usepackage{amssymb}

\usepackage{color}
\usepackage{graphicx}
\usepackage{cancel}
\usepackage{textcomp}
\usepackage{amsmath}
\usepackage{bm}
\usepackage{epsfig}
\usepackage{comment}
\usepackage{textcomp}

\graphicspath{ {./fig/} }


\begin{document}

\input{front}
\input{intro}
\input{vertex}

\input{jet}
\input{refiner}
\input{flavortag}

\input{summary}

\input{bib}

\end{document}

%% file: front.tex
% front.tex

\begin{frontmatter}

\title{LCFIPlus: A Framework for Jet Analysis in Linear Collider Studies}
\author[kyushu]{Taikan Suehara}
\ead{suehara@phys.kyushu-u.ac.jp}
\author[uticepp]{Tomohiko Tanabe}
\ead{tomohiko@icepp.s.u-tokyo.ac.jp}
\address[kyushu]{Dept. of Physics, Faculty of Sciences, Kyushu Univ., 6-10-1 Hakozaki, Higashi-ku, Fukuoka, 812-8581, Japan}
\address[uticepp]{ICEPP, The Univ. of Tokyo, 7-3-1 Hongo, Bunkyo-ku, Tokyo, 113-0033, Japan}

\begin{abstract}
We report on the progress in flavor identification tools 
developed for a future $e^+e^-$ linear collider
such as the International Linear Collider (ILC) and Compact Linear Collider (CLIC).
Building on the work carried out by the LCFIVertex collaboration,
we employ new strategies in vertex finding and jet finding,
and introduce new discriminating variables for jet flavor identification.
We present the performance of the new algorithms in the conditions
simulated using a detector concept designed for the ILC.
The algorithms have been successfully used in ILC physics simulation studies,
such as those presented in the ILC Technical Design Report.
\end{abstract}

\end{frontmatter}

%% file: intro.tex
%!TEX root = main.tex
\section{Introduction}

In high-energy collider experiments, 
the identification of the jet flavor plays an important role in event reconstruction.
The dominant decay signatures of top quarks and Higgs bosons
include bottom ($b$) jets.
Many new physics scenarios, such as those in supersymmetric theories,
have discovery signatures involving third-generation quarks.
Bottom jets can be identified by the fact that bottom quarks
form hadrons which have specific masses and prolonged lifetimes
compared to jets formed from lighter quarks.
Charged tracks resulting from bottom hadrons
tend to have sizable impact parameters
measured with respect to the primary vertex.
If there are two or more tracks originating from the same hadron,
its decay vertex can be reconstructed directly.
Bottom jets can be identified if the distance between
the primary vertex and the decay vertex is sufficiently large,
and that the reconstructed mass from the combination of the charged tracks
is consistent with that of a hadron containing a bottom quark.

More recently, the identification of the charm ($c$) jet
has also become a target of study
with the planning of future $e^+e^-$ linear colliders,
such as the International Linear Collider (ILC)~\cite{Abe:2010aa,Aihara:2009ad,Behnke:2013lya}
and Compact Linear Collider (CLIC)~\cite{Linssen:2012hp}.
The capability of $c$~jet identification allows us to probe
the complete hadronic decays of the Higgs boson
$H\rightarrow b\overline{b},\,c\overline{c},\,gg$
as well as flavor-changing neutral currents in the top sector
such as $t\rightarrow cZ$ and $t\rightarrow cH$.
Since charm hadrons have smaller masses and shorter lifetimes compared to bottom hadrons,
charm hadrons are more difficult to identify,
although the excellent point resolution in the vertex detectors
to be employed by future $e^+e^-$ linear colliders makes it possible.

An important aspect of flavor identification at high-energy $e^+e^-$ collisions
is the high multiplicity of $b$ jets resulting from
the production of multiple top quarks and Higgs bosons.
The direct measurement of the top Yukawa coupling
is to be performed in the channel containing four $b$ jets:
$e^+e^-\rightarrow t\overline{t}H\rightarrow bW^+\overline{b}W^-b\overline{b}$.
The channels most sensitive to the trilinear Higgs self-coupling contain 4--6 $b$ jets:
$e^+e^-\rightarrow ZHH\rightarrow
q\overline{q}b\overline{b}b\overline{b},\,
b\overline{b}b\overline{b}b\overline{b}$.
These processes which are produced near the kinematic threshold around $\sqrt{s}=500$~GeV 
yield jets that have significant overlap with each other.
In existing approaches, the jet finding procedure is performed 
before the vertex finding procedure.
This means that mistakes made in the association of particles into jets
are irrecoverable later, which led to lost performance.
In this work, we employ the vertex finding step before the jet finding step
to protect the secondary vertices from breaking up and grouped into different jets.

This work benefits from the existing
heavy flavor identification software, LCFIVertex~\cite{Bailey:2009ui}.
LCFIVertex has been in use in linear collider studies
presented in the ILC Letter of Intent~\cite{Abe:2010aa,Aihara:2009ad} 
and the CLIC Concept Design Report~\cite{Linssen:2012hp}.
Our software, LCFIPlus, is derived from LCFIVertex,
with enhanced features and improved algorithms.
LCFIPlus has been used in the new studies presented in the
Detector Volume of the ILC Technical Design Report~\cite{Behnke:2013lya}
by both detectors concepts for the ILC, International Large Detector (ILD)~\cite{ILDweb}
and Silicon Detector~(SiD)~\cite{SiDweb} concepts.
LCFIPlus is a modular software framework, implementing each algorithm as a module,
which can be selected and ordered, at run-time, and can use
external programs to substitute the internal algorithms.
LCFIPlus is implemented in the Marlin framework of the iLCSoft~\cite{ILCsoftweb} framework.
In essence, LCFIPlus consists of the following algorithms:
high-purity vertex finders for the reconstruction of primary and secondary vertices;
a jet finder optimized for reconstruction of heavy flavor jets;
and multivariate analysis for flavor identification based on the TMVA package~\cite{Therhaag:2009dp}.
The implementation and the performance of these algorithms 
evaluated using the full detector simulation of an ILC detector concept
are presented in the following sections.

%% file: vertex.tex
%!TEX root = main.tex
\section{Vertex Finding}

We begin our discussion with the vertex finding, which forms the basis of
our jet finding and flavor identification strategies.
As a prerequisite of the vertex finding step,
we assume that the charged tracks have been reconstructed.
In our case, we employ the {\tt MarlinTrk} framework as implemented in {\tt iLCSoft}
which interfaces a series of pattern recognition and track fitting algorithms.

Vertex finding is an essential step in discriminating the flavor of heavy quarks,
since the hadrons containing bottom or charm quarks have sizable lifetimes
which can be measured inside the detector.
The characteristic flight length,
measured in terms of the lifetime at rest $\tau$ multiplied by the speed of light $c$,
is $c\tau=400$--$500$~{\textmu m} for bottom hadrons and
$c\tau=20$--$300$~{\textmu m} for charm hadrons~\cite{PDG}.

\subsection{Vertex Fitter}

Given a set of tracks, the vertex fitter calculates the point at which the tracks are likely to have originated.
This forms the basis of the vertex finder algorithms.
We implement a simple vertex fitter for LCFIPlus as described below.

\begin{enumerate}
\item Specify an initial three-dimensional (3-D) point as follows.
\begin{enumerate}
\item If the fit has a point constraint %in addition to the track constraints
(for example, including the beam spot as a constraint for the primary vertex),
the initial value is set to the center of the point constraint.
\item In other cases, the initial value is set to the
two-dimensional (2-D) crossing point of the two circles projected from the track helical trajectories.
The $z$-direction (which is perpendicular to the projected circles)
is determined after the calculation of the 2-D point, by scanning along the helical trajectory.
\item  If the number of the tracks is three or greater, we choose the initial value of the two tracks which give the highest vertex probability.
\end{enumerate}
\item 
A 3-D fit is performed for the vertex position using the Minuit algorithm,
as implemented in the Minuit2Minizmizer class of ROOT.
We minimize the vertex $\chi^2$ value, which is constructed by adding the $\chi^2$ contribution of every track to the fitted vertex point.
(The uncertainty of the 3-D point is not considered here.)

\item The output of the vertex fitter is the minimized $\chi^2$ value as computed from the sum of the $\chi^2$ contribution from all the tracks, as well as the individual contribution to the $\chi^2$ value from each track.
The vertex uncertainty and probability are computed.

\end{enumerate}

\subsection{Primary Vertex Finder}

We describe our implementation of the primary vertex finder
as the first application of the vertex fitter.
An event containing jets has many tracks originating from the primary vertex.
We employ a tear-down algorithm for the primary vertex finder in LCFIPlus as described below.

First, the minimized vertex $\chi^2$ value is computed using all the tracks in the event.
The track which has the highest contribution to the $\chi^2$ value is removed, provided that this contribution is larger than a certain threshold value.
This process is repeated until all the remaining tracks in the vertex have contributions to the $\chi^2$ value smaller than this threshold value.
The beam spot constraint is also included in the primary vertex finder.

In the following studies, the threshold value of the track contribution to the 
$\chi^2$ value is set to 25.  The number of degrees of freedom of the fit is three.
The ILC beam size is assumed to be 
$\sigma_x = 639$ nm (horizontal), $\sigma_y = 5.7$ nm (vertical),
and $\sigma_z = 91.3$ $\mu$m (longitudinal),
following the accelerator design in the ILC Technical Design Report~\cite{Behnke:2013lya}.

\subsection{Secondary Vertex Finder}

The secondary vertex finder is used to look for the decay of $b$ and $c$ hadrons,
which are typically embedded in jets.
Conventional secondary vertex finders utilize the jet direction
in order to reduce the possible number of track combinations and therefore the required computational time.
In such a method, vertex candidates with positions that are inconsistent with the jet direction get discarded.
This leads to secondary vertices which are not found if an incorrect number of jets is specified or if there is confusion in the jet finding step, leading to inaccurate jet directions.

In our approach, we find secondary vertices without depending on the jet direction.
This approach has the benefit of increased efficiencies for the secondary vertex finding, especially in final states involving many jets.

Using the tracks which are not part of the primary vertex, 
we build all possible track pairs as seeds to the secondary vertex finder.
The $\chi^2$ value is computed with the vertex fitter for all track pairs.
Selections are applied to improve the quality of the two-track vertices based on the following criteria:
\begin{itemize}
	\item The invariant mass of the track pair must be smaller than a threshold value.
	\item The invariant mass of the track pair must be smaller than the energy of both tracks.
	\item Track pairs which pass the $V^0$ selection (described later) are discarded.
	\item The position of the reconstructed vertex with respect to the primary vertex
			must be on the same side as the sum of the track momentum.
	\item The $\chi^2$ value of the vertex must be smaller than a threshold value.
\end{itemize}
% The mass and $\chi^2$ threshold values are parameters that are specified at run-time.
We use 10 GeV for the mass and 9 for the $\chi^2$ value for the reference results.

Additional tracks are attached to the track pairs to create vertices with three or more tracks.
The addition of a track must give a $\chi^2$ contribution to the vertex smaller than a specified threshold value, and the resulting vertex must satisfy the same requirements as above on the mass, position, and $\chi^2$ value, except the $V^0$ requirement which only makes sense for track pairs.
This procedure is repeated until there are no more tracks that can be combined according to the criteria above.

The procedure so far does not preclude a track being used to form more than one vertex.
In the following step, these possible overlapping tracks are dealt with as follows.
We preferentially adopt vertices containing three or more tracks since they are less prone to combinatorial mistakes compared to two-track vertices.
The vertex with three or more tracks with the highest vertex probability is accepted as a secondary vertex.
All tracks which are used to form this vertex are eliminated from the other vertices.
The procedure is repeated until there are no more vertices with three or more tracks.
The same procedure is applied for the remaining two-track vertices, accepting in the order of highest vertex probability.
The procedure is repeated until there are no more two-track vertices to accept.
This yields a list of non-overlapping secondary vertices.

In the final step, we attempt to recover tracks in the primary vertex which may have originated from a secondary decay.
This step is optional, and is included in the reference results.
For every secondary vertex, the tracks in the primary vertex are examined based on the following two criteria.
The first criterion is that the increase in the vertex mass due to the inclusion of the track must be smaller than the track energy or the sum of the energy of all the tracks in the vertex.
The second criterion is that the track must be in the same hemisphere as the direction of the combined momenta of all the tracks in the vertex.
For tracks that satisfy the criteria above, the vertex $\chi^2$ value is computed including the track, and the track $\chi^2$ contribution to the secondary vertex is compared to that of the primary vertex.
If the track $\chi^2$ contribution to the secondary vertex is smaller than half the track $\chi^2$ contribution to the primary vertex, the track is removed from the primary vertex and attached to the secondary vertex.

\subsection{$V^0$ rejection}
\label{sec:v0}

Neutral particles which decay or convert into a pair of charged tracks,
known as $V^0$ particles,
have signatures that resemble those from the decay of bottom or charm hadrons.
It is essential to remove the $V^0$ particles in the list of particles to consider
in order to reduce the leakage originating from light jets.
Dedicated filters are applied in order to reconstruct and remove 
the following decays of $V^0$ particles:
$K_S^0\rightarrow \pi^+\pi^-$, $\Lambda^0\rightarrow p\pi^-$, and photon conversions $\gamma_\textrm{conv} X \rightarrow e^+e^- X$.

The selection of $V^0$ particles is applied to the list of secondary vertex candidates.

The identification of the $V^0$ particles is based on the kinematic variables
which are computed from the sum of the four vectors of the two charged tracks,
and the angle and distance with respect to the primary vertex.
The following variables are used:
the reconstructed mass,
the distance between the primary vertex and the $V^0$ vertex denoted as $r$,
and
the cosine of the angle between the $V^0$ momentum and 
the displacement vector from the primary vertex
to the $V^0$ vertex position denoted as $\hat{p}\cdot\hat{r}$.
The variable $\hat{p}\cdot\hat{r}$ tests
the consistency of whether the $V^0$ particle
originated from the primary vertex.

Two sets of selection criteria are used for the $V^0$ selection
as summarized in Tab.~\ref{tbl:v0cuts}.
Tracks included in the $V^0$ vertices with `tight' selection are never used as any other vertex candidates.
On the other hand, $V^0$ vertices with `loose' selection are not used as the $b$ or $c$ vertices, but
the tracks in the `loose' $V^0$ vertices are considered with other tracks to try to form other vertices.

\begin{table}[tphb]
\begin{center}
\begin{tabular}{lcccccc}
\hline
& $K_S^0$ & $K_S^0$ & $\Lambda^0$ & $\Lambda^0$ & $\gamma_\textrm{conv}$ & $\gamma_\textrm{conv}$ \\
& tight & loose & tight & loose & tight & loose \\
\hline
Mass (GeV) & $[0.493, 0.503]$ & $[0.488,0.508]$ & $[1.111,1.121]$ & $[1.106,1.126]$ & $<0.005$ & $<0.01$ \\
$r$ (mm) & $>0.5$ & $>0.3$ & $>0.5$ & $>0.3$ & $>9$ & $>9$ \\
$\hat{p}\cdot\hat{r}$ & $>0.999$ & $>0.999$ & $>0.99995$ & $>0.999$ & $>0.99995$ & $>0.999$ \\
\hline
\end{tabular}
\caption{Summary of $V^0$ selection.
Refer to the text for the definition of the variables.}
\label{tbl:v0cuts}
\end{center}
\end{table}

\subsection{Performance of vertex finding}

The performance of the vertex finders are evaluated on a sample of $b\bar{b}$ events at a center-of-mass energy of $\sqrt{s}=91.2$~GeV, including the Geant4-based full simulation of the response of the ILD detector concept.
The primary vertex finder was applied to the events after the track finding,
followed by the secondary vertex finder, using the default parameters.
Table~\ref{tbl:vtxtrk_z} gives the total number of tracks which are categorized according to their decay chain.
\begin{itemize}
\item \emph{Primary}: Tracks that originate from the primary vertex.
\item \emph{Bottom}: Tracks whose most immediate parent with a non-zero lifetime containing a bottom quark.
\item \emph{Charm}: Same as above, except the parent contains a charm quark.
\item \emph{Others}: All the other tracks, such as those from $\tau$ decays, strange hadrons, or photon conversions.
\end{itemize}
We also give the fraction of tracks that are used to form a secondary vertex.
In this sample, around 60\% of the tracks from bottom and charm hadrons are correctly used to form secondary vertices.
The fraction of tracks from the primary vertex that are erroneously used to form a secondary vertex is 0.6\%.
We give two additional measures to assess the correctness of the secondary vertex reconstruction.
First, we require that all the tracks forming the secondary vertex must be part of the same decay chain, descending from the same bottom hadron.
At this stage, we do not yet discriminate the cross-contamination between bottom and charm tracks.
We see that the secondary vertex contains very little contamination outside the decay chain, at the level of 1--2\% in terms of the number of tracks.
The second measure is more strict, requiring that all the secondary tracks must come from the same most immediate parent particle with a non-zero lifetime.
This is a measure of how well we discriminate between bottom and charm tracks: of all the tracks forming a secondary vertex, about 55--60\% of them are well separated into bottom and charm vertices.

\begin{table}[tphb] 
	\begin{center} 
		\begin{tabular}{l|c|c|c|c}
			\hline
			Track origin & Primary & Bottom & Charm & Others \\ \hline
			Total number of tracks & 496897 & 258299 & 247352 & 56432 \\ \hline
			Tracks in secondary vertices & 0.6\% & 57.5\% & 64.3\% & 2.5\% \\
			... from the same decay chain & --- & 56.6\% & 63.4\% & 1.9\% \\
			... from the same parent particle & --- & 32.2\% & 38.9\% & 1.2\% \\
			\hline
		\end{tabular}  
		\caption{The performance of the LCFIPlus vertex finder evaluated on a sample of $b\overline{b}$ events with $\sqrt{s} = 91.2$ GeV.  Refer to the main text for the explanation of the categories.}
		\label{tbl:vtxtrk_z}
	\end{center}
\end{table}

%% file: jet.tex
% jet.tex
\section{Jet Clustering}

Our jet clustering method aims to achieve the best performance of flavor tagging
at multi-jet final states, since most of interesting final states in linear colliders,
especially related to the Higgs boson, have more than 4-jets with several $b$ jets, whose
flavor tagging performance suffers from mis-jet clustering.

To avoid mis-counting of heavy jets in those processes,
we utilize variables characteristic to heavy-flavor jets, which are
existence of secondary vertices and leptons in the jets to treat as a `seed' of 
jet clustering.
This helps the heavy-flavor jets to be separated, thus is expected to result
in the better flavor tagging.

The jet finder can also run as a traditional jet finder
if the user switches off the feature to utilize secondary vertices and leptons.

\subsection{Lepton Tagging}

Isolated leptons within a jet can be a sign of
semileptonic decays of heavy flavor hadrons.
Here, we focus on muons instead of electrons,
since electron identification suffers from
the incorrect matching of calorimeter clusters with the track.
We use a simple muon selection criteria
by requiring an energy deposit of greater than 50 MeV in the muon chamber,
while limiting the energy deposits inside the
electromagnetic and hadron calorimeters.
To further increase the purity of the muon selection,
we require the impact parameter of the track
in either direction ($d_0$ or $z_0$) to be displaced from
the primary vertex by more than 5 $\sigma$.
These muons are treated in equal footing as secondary vertices
in the procedure below.

\subsection{Vertex and lepton combination}

A striking feature of heavy flavor hadrons is the cascade of multiple decays.
The purpose of this step is to combine
the secondary vertices and the leptons from the semileptonic decays
in a way that is consistent with the cascade decay.
The combination is done using the opening angles between the vertices and/or leptons.
For the vertex, the direction of the vertex position from the primary vertex is used,
while for the leptons the momentum direction is used.
A pair of two vertices are combined if the opening angle between the two vertices is
less than 0.2~rad.
For a pair of two leptons or a lepton and a vertex,
the opening angle threshold is 0.3~rad,
considering the fact that leptons tend to have a larger deviation in angle
with respect to the jet direction.

\subsection{Jet clustering}

The jet clustering is the last step of our method.
First, the vertices and leptons are treated as jet cores.
If the number of jet cores is larger than the required number of jets,
the nearest jet cores are combined until the required number is reached.
The resulting jet cores are kept separate in the procedure below.

Second, the remaining tracks and neutral clusters,
including those that come from the primary,
are combined to one of the jet cores.
We perform this in two steps, first with a cone jet clustering algorithm
and then with the traditional Durham\cite{Catani:1991hj}-like clustering algorithm.
By looking at the opening angle between the momentum direction of the particle
and that of the jet cores,
those which fall within 0.2 radian of the jet core are merged with that jet core.
If there are multiple possible jet cores to combine,
the one with the closest jet core is used.

The remaining particles (tracks or clusters) are
combined to the jet cores based on the modified Durham distance measure of
\begin{equation}
Y(i,j) = \frac{2\,\mathrm{min}\left(E_i, E_j\right)^2(1-\cos\theta_{ij})}{Q^2} + \alpha,
\end{equation}
where $E_i$ and $E_j$ stand for the jet energies,
and $\theta_{ij}$ is the angle between the two jets.
The specific energy, which is constant for all events, is given as $Q^2$,
which is typically the center-of-mass energy.
$\alpha$ is a constant to be added in the normal Durham $Y(i,j)$,
to be used to prevent the jet cores from merging with each other.
$\alpha$ is $100$ if both jets have cores and $0$ for others.
The $\alpha$ with two jet cores are tunable, so if set to $0$
there is no tendency to separate two jet cores which gives
very similar results to traditional Durham jet clustering.

Performance of the jet clustering has been studied elsewhere~\cite{Suehara:2011sq}.

The jets reconstructed by this algorithm include vertex information.
If users use an external algorithm for the jet clustering instead of this algorithm,
the vertices should be attached to each jet prior to the next step.
This can be optionally done at the beginning of the jet vertex refiner.

%% file: refiner.tex
% refiner.tex

\section{Jet Vertex Refiner}

The vertex finder in LCFIPlus does not initially rely on jets.
After the jet clustering step, the additional information about jets is used to improve the vertex reconstruction and the flavor tagging performance.
The purpose of the vertex refining step is to improve the separation between $b$ and $c$ jets by correctly reconstructing the number of secondary vertices as much as possible.
The presence of two vertices strongly indicates that the jet is a $b$ jet.
We wish to increase the efficiency of finding $b$ jets with two vertices while reducing the contamination from $c$ jets.

The refining procedure has two steps: (a) single track vertex finder, and (b) vertex combiner.
The single track vertex finder is designed to look for $b$ jets containing only one reconstructed secondary vertex with a possible track that could be interpreted as the result of an additional secondary decay.
The cascade decays in a $b$ jet are expected to result in decay points that are nearly collinear with the primary vertex.  If only one secondary vertex is found, and if there is a track whose trajectory passes near a point collinear to the primary and secondary vertices, then the track is taken as a pseudo-vertex.
We adopt a track as a single-track pseudo-vertex if it satisfies the following criteria:
\begin{itemize}
	\item The track has an impact parameter significance larger than $5 \sigma$ in either the transverse ($d_0$) or longitudinal ($z_0$) directions.
	\item The opening angle between the track trajectory and the \emph{vertex line}, defined as the line passing through the primary vertex to the secondary vertex, is less than 0.5~rad at the track's closest approach to the vertex line.
	\item The displacement from the primary vertex to the \emph{vertex point}, defined as the point along the track trajectory closest to the vertex line, and the track direction can be put into the same hemisphere.
	\item The distance from the primary vertex to the vertex point is between 0.3 and 30~mm.
	\item The distance between the vertex point and the vertex line is more than 10 times smaller than the distance from the primary vertex to the vertex point.
\end{itemize}

The vertex combiner is the next step in the vertex refining procedure.
The purpose is to reduce the number of vertices down to at most two by combining neighboring vertices.
Here, we ignore cases where the jet could rightly contain more than two secondary vertices, such as the decay of $B_c$ mesons.
The single-track pseudo-vertex is the first target for combining with other vertices, since the track often turns out to be a part of the existing secondary vertex.
The vertex combiner works as follows:
\begin{itemize}
	\item Choose a pair of vertices in a jet as \emph{seed vertices}.
	\item List all the tracks in the jet that are neither part of the primary vertex nor the seed vertices.
	\item Test each track by attaching it to each of the seed vertices. Adopt the track-vertex combination which gives a smaller track $\chi^2$ contribution in the refitted vertex.
	\item After all the tracks are attached, compute the sum of the two vertex $\chi^2$ values.
	\item Repeat with all possible pairs of vertices. Adopt the result that gives the smallest $\chi^2$ sum.
\end{itemize}
This combination step is followed by an optimization step, which works as follows:
\begin{itemize}
	\item List all the tracks from the two vertices in the jet.
	\item Test each track by removing it from its original vertex and attaching it to the other vertex.
	\item Adopt the new track assignment if the track $\chi^2$ contribution to the other vertex is smaller than that to the original vertex.
	\item Iterate once over all the tracks in the two vertices. 
	\item Compare the sum of the $\chi^2$ values of the original two vertices against that of the track-optimized vertices. If the new $\chi^2$ sum is smaller, the new vertices are adopted in place of the original ones.
	\item Repeat from the beginning if the vertices are updated, up to three times.
\end{itemize}

The final procedure of the vertex refiner is to select the vertices after the combination.
First, the loose $V^0$ rejection is applied as described in Sec.~\ref{sec:v0}.
Second, if the jet has two vertices, all the tracks from the two vertices are tried to be combined into a single vertex. If the probability of the combined vertex is larger than $10^{-3}$, the two vertices are replaced by the combined vertex.
The final selection is to check the distance between the two secondary vertices.
If the two vertices are separated too far, the second vertex becomes less likely to have come from a charm hadron in the cascade decay.
We drop the second vertex if the distance divided by the jet energy is greater than $0.1$~mm\,$\cdot$\,GeV$^{-1}$.

\begin{table}[p]
	\begin{center}
		\begin{tabular}{c|c|c|c}\hline\hline
			(\#vtx, \#pseudo-vtx) &    $b$ jet &    $c$ jet &    $uds$ jet \\ \hline
			$(0,0)$ & 21.3\%  & 59.3\%  & 98.1\%  \\
			$(0,1)$ &  1.61\% &  0.17\% &  0.01\% \\
			$(1,0)$ & 39.7\%  & 39.8\%  &  1.80\% \\
			$(1,1)$ & 13.5\%  &  0.54\% &  0.02\% \\
			$(2,0)$ & 23.8\%  &  0.19\% &  0.04\% \\ \hline\hline
		\end{tabular}
		\caption{The distribution of $b$, $c$, and $uds$ jets categorized in terms of the reconstructed number of vertices and single-track pseudo-vertices, studied in a sample of $e^+e^-\rightarrow q\overline{q}$ events at $\sqrt{s}=91$~GeV.}
		\label{tbl:refvtx}
	\end{center}
\end{table}

Table~\ref{tbl:refvtx} shows the result of the jet vertex refiner evaluated in terms of number of vertices, using samples of $e^+e^-\rightarrow b\overline{b}$, $c\overline{c}$, and $q\overline{q}$, where $q$ is any of the lighter $uds$ quark, at the center-of-mass energy of $\sqrt{s} = 91$ GeV.
The events are forced into two jets.  The jets are categorized and counted according to the number of vertices (\#vtx) and pseudo-vertices (\#pseudo-vtx).
In the category of two vertices, 24\% of $b$ jets fall into this category, while $c$ jets and $uds$ jets are highly suppressed.
In the category of one vertex and one pseudo-vertex, we succeed in recovering 13.5\% of the $b$ jets, which otherwise would have been grouped together in the one vertex category, with low contamination coming from the lighter jets.
The $uds$ jets are confined very well in the zero vertex category (98.1\%), which shows a good separation of $uds$ jets from the $b$ and $c$ jets.

%% file: flavortag.tex
% flavortag.tex
\section{Flavor Tagging}

While the flavor tagging procedure described in this section is in principle
independent of the order of the jet finding and secondary vertex finding.
the performance is given for the procedures given in this paper,
i.e. in the order of vertex finding, jet finding, and vertex refining
as described in the previous sections.

The flavor tagging procedure is based on a multivariate classifier
as implemented in the TMVA package.
The flavor tagging procedure is applied to each jet and
makes no attempt to look at the interaction between the jets
beyond what is implemented up to this point.
The jets are divided into four categories according
the number of reconstructed vertices in a jet.
For each category, a set of input variables are defined,
which are then passed to the multivariate classifier.
The classifier response is normalized across the different categories,
which can then be used in a physics analysis.

We employ boosted decision trees (BDTs) as the multivariate classifier
in the TMVA package in ROOT.  The BDTs with gradient boosting are used.
The BDTs operate in the multiclass mode which allows the
simultaneous training of multiple classes of events.
In our case, we define three classes, which are $b$ jets, $c$ jets, and $uds$ jets.

The jets are categorized by the number of reconstructed vertices.
By the design of the vertex refiner described in the previous section,
each jet can either have zero, one, or two properly reconstructed vertices.
In addition, each jet can have he single-track pseudovertex is also considered.
We separate the jets into the four categories as listed in
Tab.~\ref{tab:flavtag_categories}
\begin{table}
\centering
\begin{tabular}{lcccc}
\hline\hline
Category             &  A & B & C & D \\
\hline
Number of vertices   & 0 & 1 & 1 & 2 \\
Number of single-track pseudovertices & 0-2 & 0 & 1 & 0 \\
\hline\hline
\end{tabular}
\caption{Definition of flavor tagging categories.}
\label{tab:flavtag_categories}
\end{table}
  
The flavor tagging input variables are constructed from the constituents of the jets
such as the charged tracks and secondary vertices.
The momentum of the jet itself is used for
the inspection of the jet constituents in terms of the jet direction.
Many input variables can depend on the energy of the jet,
since the decay length and angles between particles
necessarily depend on the boost of the particles involved.
They can be normalized making use of the jet energy to diminish the jet energy dependence.
The jet energy dependence cannot be completely eliminated because
the acceptance cuts and the detector effects
are inherently not invariant as a function of the jet energy.
The list of input variables are shown in Tabs.~\ref{tab:flavtag_variables}-\ref{tab:flavtag_variables2}.

\begin{figure}[htbp]
\centering
\begin{minipage}{0.48\linewidth}
\includegraphics[width=1\linewidth]{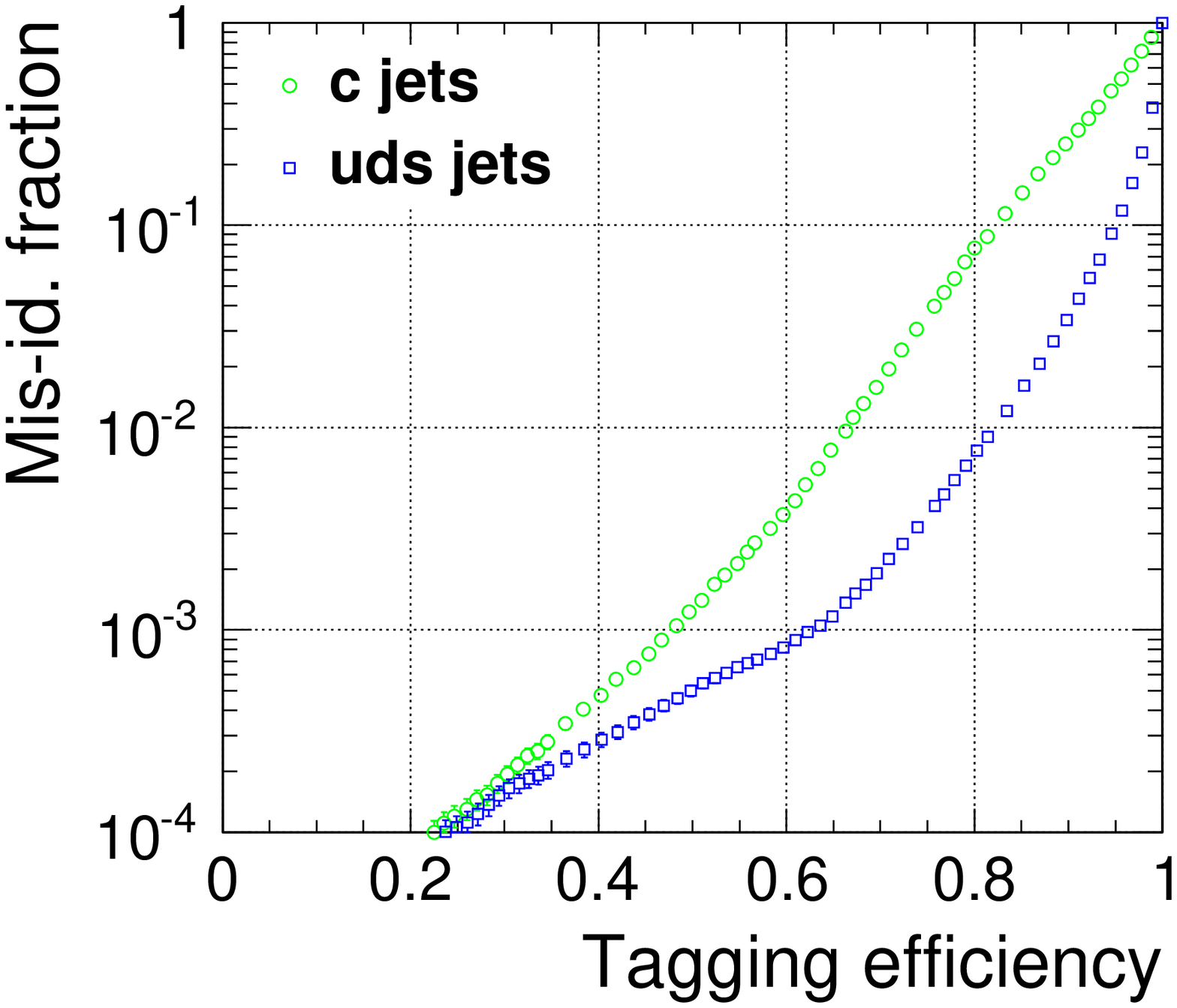}
\centering
(a) $b$ tag
\end{minipage}
\begin{minipage}{0.48\linewidth}
\includegraphics[width=1\linewidth]{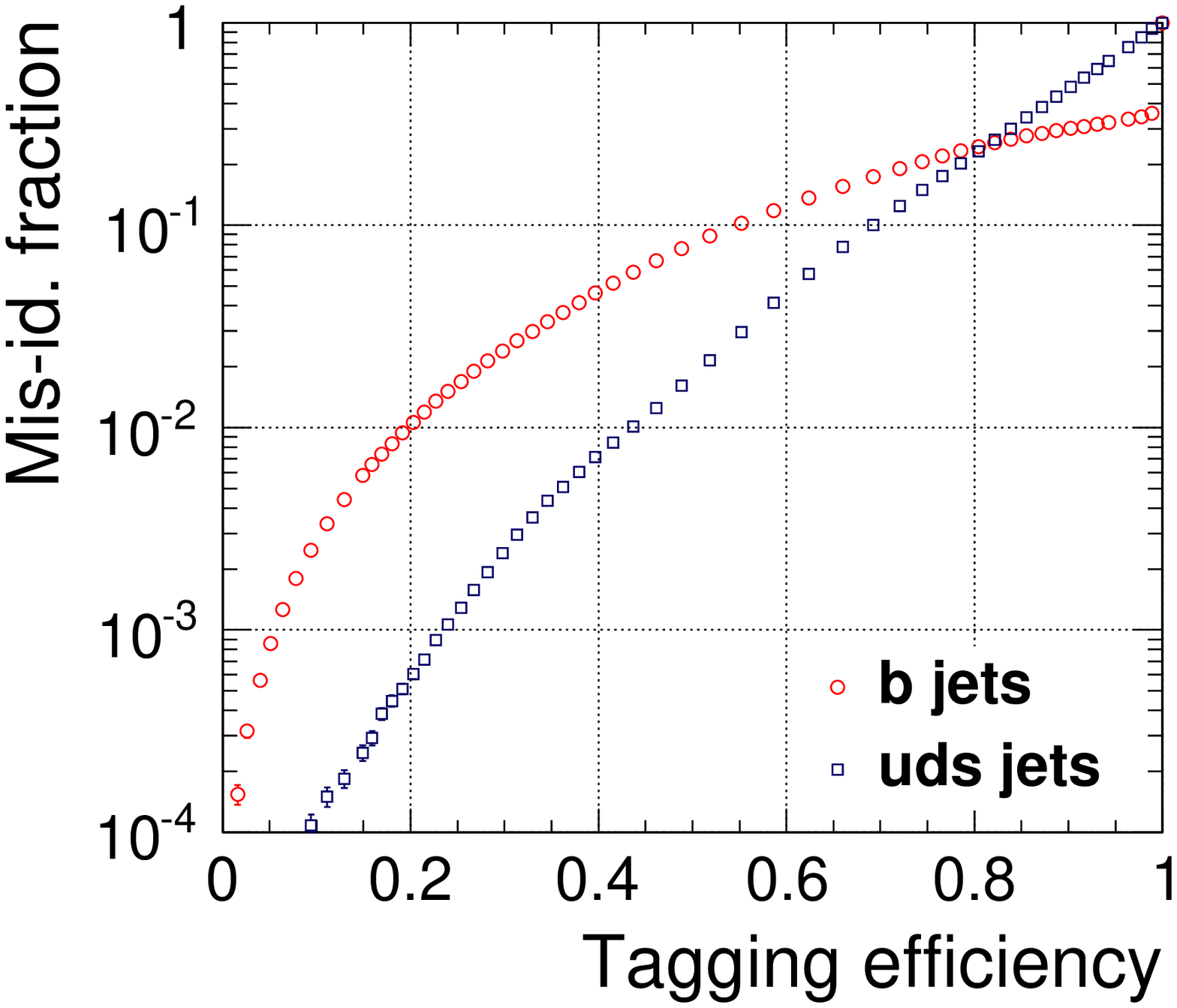}
\centering
(b) $c$ tag
\end{minipage}
\caption{The flavor tagging performance, evaluated on $Z\rightarrow q\overline{q}$ sample at $\sqrt{s}=91.2$~GeV,
is shown in terms of the mis-identification fraction versus the tagging efficiency.
(a) The tagging efficiency is shown for $b$ jets.
The green (circle) points show the fraction of $c$ jets being mistaken as a $b$ jet.
The blue (square) points show the fraction of $uds$ jets being mistaken as a $b$ jet.
(b) The tagging efficiency is shown for $c$ jets.
The red (circle) points show the fraction of $b$ jets being mistaken as a $c$ jet.
The blue (square) points show the fraction of $uds$ jets being mistaken as a $c$ jet.
}
\label{fig:flavtag-zpole}
\end{figure}

\begin{figure}[htbp]
\centering
\begin{minipage}{0.48\linewidth}
\includegraphics[width=1\linewidth]{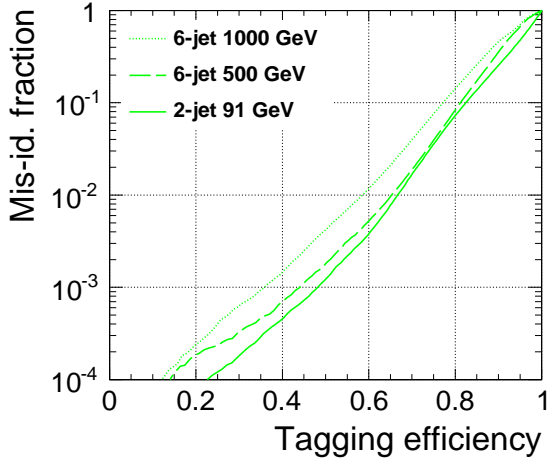}
\centering
(a) $b$ tag with $c$ background
\end{minipage}
\begin{minipage}{0.48\linewidth}
\includegraphics[width=1\linewidth]{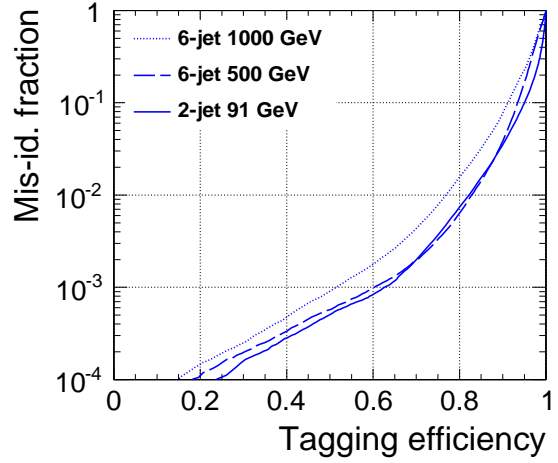}
\centering
(b) $b$ tag with $uds$ background
\end{minipage}
\begin{minipage}{0.48\linewidth}
\includegraphics[width=1\linewidth]{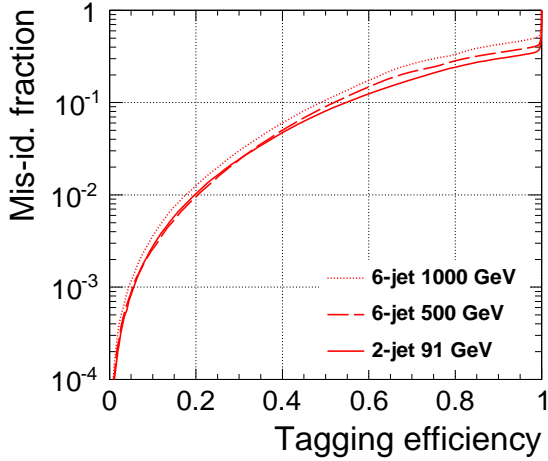}
\centering
(c) $c$ tag with $b$ background
\end{minipage}
\begin{minipage}{0.48\linewidth}
\includegraphics[width=1\linewidth]{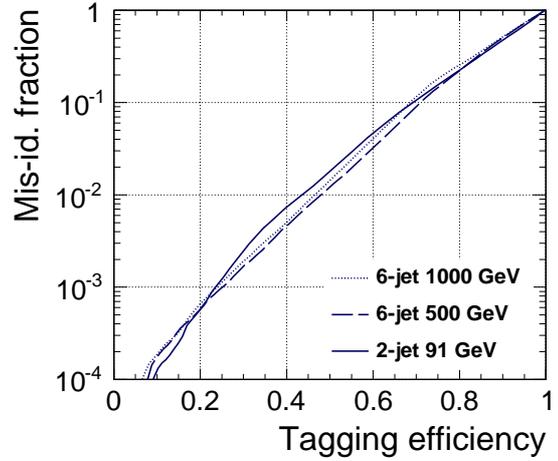}
\centering
(d) $c$ tag with $uds$ background
\end{minipage}
\caption{Comparison of flavor tagging performance on processes using training results of each processes.
Solid lines correspond to the case of 2-jet events at $\sqrt{s}=91$~GeV.
Dashed lines correspond to the case of 6-jet events at $\sqrt{s}=500$~GeV.
Dotted lines correspond to the case of 6-jet events at $\sqrt{s}=1000$~GeV.
(a) The tagging efficiency is shown for $b$ jets with the mis-identification fraction evaluated using $c$ jets.
(b) The tagging efficiency is shown for $b$ jets with the mis-identification fraction evaluated using $uds$ jets.
(c) The tagging efficiency is shown for $c$ jets with the mis-identification fraction evaluated using $b$ jets.
(d) The tagging efficiency is shown for $c$ jets with the mis-identification fraction evaluated using $uds$ jets.
}
\label{fig:flavtag-comparison}
\end{figure}

The performance of the flavor tagging for
$e^+e^-\rightarrow Z\rightarrow q\overline{q}$ two jet samples
for $\sqrt{s}=91.2$~GeV
is shown in Fig.~\ref{fig:flavtag-zpole}.
Two plots are shown corresponding to the
performance of the $b$ tagging and $c$ tagging.
The two lines correspond to the difference sources of misidentification.
For the $b$ tagging performance, we show separately the fraction of $c$ jets and $uds$ jets
that are misidentified as $b$ jets.
Similarly for the $c$ tagging performance, we show the leakage from $b$ jets and $uds$ jets.
For a fixed $b$ tagging efficiency at $\epsilon_b=80\%~(50\%)$,
the misidentification fraction is 7.3\% (0.12\%) and 0.74\% (0.051\%)
for $c$ jets and $uds$ jets, respectively.
The rate of $c$ jets misidentified as a $b$ jet
is larger than that for $uds$ jets as expected
due to the lifetime of the charm hadrons.
For a fixed $c$ tagging efficiency at $\epsilon_c=80\%~(50\%)$,
the misidentification fraction is 22\% (8.1\%) and 24\% (1.8\%)
for $b$ jets and $uds$ jets, respectively.
The contamination in $c$ tagging is higher than that for
$b$ tagging since both $b$ jets and $uds$ can both resemble $c$ jets.
For the high efficiency $c$ tagging region,
the contamination from $b$ jets drops sharply,
since $b$ jets with two identified secondary vertices
can be discriminated easily from $c$ jets,
which can only have up to one secondary vertex.
If the $b$ jet contains only one secondary vertex,
the discrimination against $c$ jet become challenging
as seen in the relatively large leakage rate in the low efficiency region.

Figure \ref{fig:flavtag-comparison} shows comparison plots with different event samples.
In addition to $q\overline{q}$ 91 GeV sample, 1000 and 500 GeV 6-jet samples are employed.
For 6-jet samples,
$6b$, $6c$ and combination of $6u$, $6d$ and $6s$ events mainly from $ZZZ$ production
are used as the source of $b$, $c$ and $uds$ jets with fixed 6-jet clustering in LCFIPlus,
Training of TMVA is done with the same final states as the test samples.
Clear degradation at 1000 GeV $6q$ events are seen at $b$-tagging and
$c$-tagging with $b$-background at high tagging efficiency region.
Degradation at 500 GeV $6q$ events are much smaller, showing
robustness of LCFIPlus flavor tagging performance in the dense jet environments.

\begin{table}
\centering
\scalebox{0.8}{%
\begin{tabular}{lp{10cm}p{2cm}p{2cm}}
\hline\hline
Name & Description & Normalization factor & Used by category \\
\hline
trk1d0sig
&  d0 significance of track with highest d0 significance
& $1$
& A, B, C, D
\\
trk2d0sig
& d0 significance of track with second highest d0 significance 
& $1$
& A, B, C, D
\\
trk1z0sig
& z0 significance of track with highest d0 significance% (ordering by d0, not z0) 
& $1$
& A, B, C, D
\\
trk2z0sig
& z0 significance of track with second highest d0 significance% (ordering by d0, not z0) 
& $1$
& A, B, C, D
\\
trk1pt      
& transverse momentum of track with highest d0 significance
& $1/E_\mathrm{jet}$
& A, B, C, D
\\
trk2pt      
& transverse momentum of track with second highest d0 significance
& $1/E_\mathrm{jet}$
& A, B, C, D
\\
jprobr      
& joint probability in the r-phi plane using all tracks
& $1$
& A, B, C, D
\\
jprobr5sigma
& joint probability in the r-phi plane using all tracks having impact parameter significance exceeding 5 sigma
& $1$
& A, B, C, D
\\
jprobz      
& joint probability in the z projection using all tracks
& $1$
& A, B, C, D
\\
jprobz5sigma
& joint probability in the z projection using all tracks having impact parameter significance exceeding 5 sigma
& $1$
& A, B, C, D
\\
d0bprob     
& product of b-quark probabilities of d0 values for all tracks, using b/c/q d0 distributions
& $1$
& A, B, C, D
\\
d0cprob     
& product of c-quark probabilities of d0 values for all tracks, using b/c/q d0 distributions
& $1$
& A, B, C, D
\\
d0qprob     
& product of q-quark probabilities of d0 values for all tracks, using b/c/q d0 distributions
& $1$
& A, B, C, D
\\
z0bprob     
& product of b-quark probabilities of z0 values for all tracks, using b/c/q z0 distributions
& $1$
& A, B, C, D
\\
z0cprob     
& product of c-quark probabilities of z0 values for all tracks, using b/c/q z0 distributions
& $1$
& A, B, C, D
\\
z0qprob     
& product of q-quark probabilities of z0 values for all tracks, using b/c/q z0 distributions
& $1$
& A, B, C, D
\\
nmuon	    
& number of identified muons
& $1$
& A, B, C, D
\\
nelectron   
& number of identified electrons
& $1$
& A, B, C, D
\\
trkmass     
& mass of all tracks exceeding 5 sigma significance in d0/z0 values
& $1$
& A, B, C, D
\\
\hline\hline
\end{tabular}
}
\caption{Flavor tagging input variables.
The category is defined in Tab.~\ref{tab:flavtag_categories}.
}
\label{tab:flavtag_variables}
\end{table}

\begin{table}
\centering
\scalebox{0.8}{%
\begin{tabular}{lp{10cm}p{2cm}p{2cm}}
\hline\hline
Name & Description & Normalization factor & Used by category \\
\hline
1vtxprob    
& vertex probability with all tracks associated in vertices combined
& $1$
& B, C, D
\\
vtxlen1     
& decay length of the first vertex in the jet (zero if no vertex is found)
& $1/E_\mathrm{jet}$
& B, C, D
\\
vtxlen2     
& decay length of the second vertex in the jet (zero if number of vertex is less than two)
& $1/E_\mathrm{jet}$
& D
\\
vtxlen12    
& distance between the first and second vertex (zero if number of vertex is less than two)
& $1/E_\mathrm{jet}$
& D
\\
vtxsig1     
& decay length significance of the first vertex in the jet (zero if no vertex is found)
& $1/E_\mathrm{jet}$
& B, C, D
\\
vtxsig2
&  decay length significance of the second vertex in the jet (zero if number of vertex is less than two)
& $1/E_\mathrm{jet}$
& D
\\
vtxsig12  
& vtxlen12 divided by its error as computed from the sum of the covariance matrix of the first and second vertices, projected along the line connecting the two vertices
& $1/E_\mathrm{jet}$
& D
\\
vtxdirang1   
&   the angle between the momentum (computed as a vector sum of track momenta) and the displacement of the first vertex
& $E_\mathrm{jet}$
& B, C, D
\\
vtxdirang2     
&  the angle between the momentum (computed as a vector sum of track momenta) and the displacement of the second vertex
& $E_\mathrm{jet}$
& D
\\
vtxmult1       
&  number of tracks included in the first vertex (zero if no vertex is found)
& $1$
& B, C, D
\\
vtxmult2       
&  number of tracks included in the second vertex (zero if number of vertex is less than two)
& $1$
& D
\\
vtxmult        
&  number of tracks which are used to form secondary vertices (summed for all vertices)
& $1$
& D
\\
vtxmom1        
&  magnitude of the vector sum of the momenta of all tracks combined into the first vertex
& $1/E_\mathrm{jet}$
& B, C, D
\\
vtxmom2        
&  magnitude of the vector sum of the momenta of all tracks combined into the second vertex
& $1/E_\mathrm{jet}$
& D
\\
vtxmass1       
&  mass of the first vertex computed from the sum of track four-momenta
& $1$
& B, C, D
\\
vtxmass2       
&  mass of the second vertex computed from the sum of track four-momenta
& $1$
& D
\\
vtxmass        
&  vertex mass as computed from the sum of four momenta of all tracks forming secondary vertices
& $1$
& B, C, D
\\
vtxmasspc      
&  mass of the vertex with minimum pt correction allowed by the error matrices of the primary and secondary vertices
& $1$
& B, C, D
\\
vtxprob        
&  vertex probability; for multiple vertices, the probability P is computed as 1-P = (1-P1)(1-P2)...(1-PN)
& $1$
& B, C, D
\\
\hline\hline
\end{tabular}
}
\caption{Flavor tagging input variables (continued).}
\label{tab:flavtag_variables2}
\end{table}

%% file: summary.tex
% summary.tex
\section{Summary}

We developed a set of software tools for jet clustering and flavor tagging for future linear colliders.
It includes a high-purity secondary vertex finder based on build-up vertex clustering,
a jet clustering algorithm using vertex information, and multivariate jet flavor
tagging for the separation of $b$ and $c$ jets.
The performance has been demonstrated to meet the ILC physics goals.
Improvements over previous flavor tagging software are seen in particular
in 6-jet final states, which is an important signature for the measurement
of the Higgs self-coupling. Further improvements are foreseen, such as
the implementation of a lepton finder, and further improvements of 
beam-related background reduction.

\section*{Acknowledgments}

The authors wish to thank Junping Tian, Steve Aplin and Jan Strube for the supporting work.
They are grateful to the ILC physics study groups and CLIC detector and physics study group for providing useful feedback.
The computing resource was provided in large part by the KEK computing center.
This work was partly supported by JSPS KAKENHI Grant Number 23000002.

%% file: bib.tex
% bib.tex
%%%%%%%%%%%%%%%%%%%%%%%
%% Elsevier bibliography styles
%%%%%%%%%%%%%%%%%%%%%%%
%% To change the style, put a % in front of the second line of the current style and
%% remove the % from the second line of the style you would like to use.
%%%%%%%%%%%%%%%%%%%%%%%

%% Numbered

%% Numbered without titles
%\bibliographystyle{model1a-num-names}

%% Harvard
%\bibliographystyle{model2-names.bst}\biboptions{authoryear}

%% Vancouver numbered
%\usepackage{numcompress}\bibliographystyle{model3-num-names}

%% Vancouver name/year
%\usepackage{numcompress}\bibliographystyle{model4-names}\biboptions{authoryear}

%% APA style
%\bibliographystyle{model5-names}\biboptions{authoryear}

%% AMA style
%\usepackage{numcompress}\bibliographystyle{model6-num-names}

%% `Elsevier LaTeX' style
\bibliographystyle{elsarticle-num}
%%%%%%%%%%%%%%%%%%%%%%%

\bibliography{main}